\documentclass[epj-spec]{svjour} 

\usepackage{graphicx}
\usepackage{latexsym}
\usepackage{amssymb}
\usepackage{amsmath}
\begin{document}
\title{The effects of forcing and dissipation on phase transitions in
  thin granular layers.}
\author{Alexander E. Lobkovsky\inst{1}\and Francisco Vega Reyes\inst{2} \and J. S. Urbach\inst{3} \fnmsep\thanks{\email{urbach@physics.georgetown.edu}} }
\institute{National Institutes of Health, Bethesda, MD 20894 \and
  Departamento de F\'{\i}sica, Universidad de Extremadura, E-06071
  Badajoz, Spain \and Department of Physics, Georgetown University,
  Washington DC, 20057}
\abstract{Recent experimental and computational studies of vibrated
  thin layers of identical spheres have shown transitions to ordered
  phases similar to those seen in equilibrium systems. Motivated by
  these results, we carry out simulations of hard inelastic spheres
  forced by homogenous white noise. We find a transition to an ordered
  state of the same symmetry as that seen in the experiments, but the
  clear phase separation observed in the vibrated system is
  absent. Simulations of purely elastic spheres also show no evidence
  for phase separation.  We show that the energy injection in the
  vibrated system is dramatically different in the different phases,
  and suggest that this creates an effective surface tension not
  present in the equilibrium or randomly forced systems. We do find,
  however, that inelasticity suppresses the onset of the ordered phase
  with random forcing, as is observed in the vibrating system, and that the amount of the suppression is proportional to the degree of inelasticity.  The
  suppression depends on the details of the energy injection mechanism, but is completely eliminated when inelastic collisions are replaced by uniform system-wide energy dissipation. 
} 
\maketitle
\section{Introduction}
\label{intro}

Excited granular materials show a variety of interesting
non-equilibrium phenomena that raise important fundamental and
practical questions \cite{aranson06}.  Studies on steady states of
simple model systems comprised of identical, spherical, non-cohesive
inelastic spheres have proven to be valuable for developing and
testing extensions of statistical mechanics to non-equilibrium
systems.  Measurements of velocity distributions have revealed
significant deviations from equilibrium Maxwellian distribution
function
\cite{olafsen98,olafsen99,losert99,baxter03,reis07,baxter07,burdeau09},
and velocity correlations show violations of molecular chaos
\cite{prevost02}. These observations are reasonably well described by
non-equilibrium kinetic theory, which shows that non-equilibrium
effects arise from the mechanism of energy injection
\cite{vannoije98,vannoije99,pagonabarraga02}.

While the `microscopics' of inelastic particles differs significantly
from their equilibrium analogs, the macroscopic phase behavior of
non-equilibrium steady states of excited granular media can show
remarkable similarity to equilibrium systems.  The crystallization of
a single layer of vibrated spheres into a hexagonally ordered array
can be directly mapped onto the analogous 2D phase transition
\cite{olafsen05,reis06}.  In the presence of a confining lid, we have
reported a complex phase diagram that is closely related to that
observed in similarly confined equilibrium colloidal systems,
including two-layer crystals with square or hexagonal symmetry
\cite{prevost04,melby05,reyes08}. In a recent study Clerc et
al. \cite{clerc08} extended this work to quasi-one-dimensional
systems, and showed that the transition to an ordered phase was
mediated by traveling waves and was triggered by negative
compressibility. However, in spite of the strong similarities between
the phase diagrams of equilibrium and granular systems, experimental
observations indicate that phase transitions in granular systems
exhibit a richer and more complex behavior.  There are ordered phases
that arise solely from nonequilibrium effects, such as the hexagonal
collapse observed in granular monolayers at low vibration amplitudes
\cite{olafsen98}.  Experimental and computational studies have shown
that inelasticity suppresses the transition to the ordered phase
\cite{reyes08,clerc08}, but the mechanism of this suppression is still
not well understood.

Motivated by these results, we have performed computational studies of
identical, spherical, inelastic spheres in a quasi-2D geometry,
excited either by random forces or by external vibration.  The
transition to the ordered phase in the randomly driven system is
suppressed when inelasticity is increased, as in the vibrated system.
However, unlike spheres under vibration, there is no clear phase
separation in the case of random driving.  We show that the energy
injection in the vibrated system is strongly phase-dependent, and
suggest that this dependence creates an effective surface tension that
accounts for the strong phase separation observed under vibration.  In
addition, we show that the suppression of the ordered phase in the
randomly driven system depends in a non-trivial way on both the form
of the energy injection and the form of the dissipation.

\section{Computational methods}
We report results from two simulation techniques: soft sphere
molecular dynamics used to simulate the experiments with external
vibration, and hard sphere event driven simulations with random energy
injection.  In all cases we simulate thin granular layers, with
identical spherical particles in a three dimensional domain that has
periodic boundary conditions in two dimensions (simulating the
horizontal plane in the vibrating plate experiments).  In the third
(`vertical') dimension, the spheres are confined to a space of
$1.75\sigma$, where $\sigma$ is the sphere diameter.  A constant
gravitational acceleration is included in the simulations of vertical
vibrations, but not the random driving simulations.

\subsection{Soft sphere molecular dynamics simulations of vertical
  vibrations}
We simulate the vertical vibration of the experiments by imposing a
sinusoidal oscillation on the horizontal boundaries with amplitude $A$
and frequency $\nu$ (keeping the spacing between the boundaries
fixed).  We have used a collisional model of rough (the particles have
rotational kinetic energy), soft (there is particle-particle overlap
during collisions) inelastic (collisions are dissipative) spheres. The
collisions are characterized by three forces. Two forces are normal to
the direction of collision, one is an elastic restoring force
$\mathbf{F}^{\mathrm{rest}}$, proportional to particle/particle
overlap, and the other is a frictional dissipative force
$\mathbf{F}^{\mathrm{diss}}$, proportional to relative normal
velocity. The third force $\mathbf{F}^{\mathrm{shear}}$, which is
tangential, is also frictional and dissipative (inelastic). The same
types of forces are used for particle-wall collisions (there is also
particle-wall overlap during these
collisions). 
The interactions included in the model do not capture the full
complexity of real inelastic collisions \cite{louge94}, but do
reproduce the dominant effects of vibration, collisions, and
dissipation. Furthermore, this collisional model has successfully
reproduced a wide variety of phenomena that are observed in
experiments, including pattern formation \cite{rapaport04}, clustering
\cite{nie00}, rheological behavior \cite{silbert01}, impurity
segregation \cite{sun06}, velocity correlations \cite{prevost02},
depletion forces \cite{melby07} and phase transitions
\cite{prevost04,melby05}. Moreover, simulations with this collisional
model reproduce our experimental observations that ordered phases are
suppressed in highly inelastic materials \cite{reyes08}.

The wall-particle forces can be expressed as

\begin{equation}
  \mathbf{F}_{iw}^{rest}=Y_wm_i(|\mathbf{r}_{iw}|-\sigma)\mathbf{\hat{r}}_{iw}, \quad\quad 
  \mathbf{F}_{iw}^{diss}=-\gamma_{nw}m_i\mathbf{v}_{iw}^n, \quad\quad
  \mathbf{F}_{iw}^{shear}=-\gamma_{sw}m_i\mathbf{v}_{iw}^t. \label{forcesMD}
\end{equation} The particle-particle interactions have analogous forms

\begin{equation}
  \mathbf{F}_{ij}^{rest}=Ym_i(|\mathbf{r}_{ij}|-\sigma)\mathbf{\hat{r}}_{ij}, \quad\quad
  \mathbf{F}_{ij}^{diss}=-\gamma_{n}m_i\mathbf{v}_{ij}^n, \quad\quad
  \mathbf{F}_{ij}^{shear}=-\gamma_{s}m_i\mathbf{v}_{ij}^t, \label{particleforcesMD}
\end{equation} 
In the above equations (\ref{forcesMD}, \ref{particleforcesMD}),
subscripts $a,b$ may be $w$, which stands for the walls, or $i,j$,
which stand for particles; $m_i$ is the particle mass;
$\mathbf{r}_{ab}$ are relative positions:
$\mathbf{r}_{ab}=\mathbf{r}_a-\mathbf{r}_b$, and $\mathbf{\hat
  r}_{ab}=\mathbf{r}_{ab}/|\mathbf{r}_{ab}|$.  Analogously,
$\mathbf{v}_{ab}$ stands for relative velocities, and
$\mathbf{v}_{ab}^n$, $\mathbf{v}_{ab}^t$ stand for the projections of
relative velocities in the normal and tangential directions
respectively. The coefficient $Y$ is the Young modulus that
characterizes the restoring force whereas $\gamma_n$, $\gamma_s$
account for the dissipation in the normal and tangential directions
respectively. In this work we have used values of parameters that
mimic the behavior of metallic balls with a coefficient of normal
restitution $\alpha\sim0.87$ \cite{sun06}, which is an intermediate
value between the experimental values $\alpha=0.95$ and $\alpha=0.77$
for steel and brass spheres \cite{reyes08}. We have used the same
values of force parameters for the wall-particle
interactions. Specifically, we use $Y=Y_w=10^7$~s$^{-2}$,
$\gamma_n=\gamma_{nw}=200$~s$^{-1}$,
$\gamma_s=\gamma_{sw}=200$~s$^{-1}$.

In order to calculate the energy input from the vibrating plate, we
define the magnitude
\begin{equation}
  W_{in}=\left|\Sigma_i ^{bin}(\mathbf{F}_{iw} \cdot \mathbf{v}_idt+(\mathbf{r}_{iw}\times\mathbf{F}_{iw})\cdot\frac{d\boldsymbol{\omega}_i}{dt}dt)\right| \label{workinput}
\end{equation}
(with $dt$ the time step in the simulation, $\boldsymbol{\omega}_i$
the particle angular velocity, and
$\mathbf{F}_{iw}=\mathbf{F}_{iw}^{\mathrm{rest}}+\mathbf{F}_{iw}^{\mathrm{diss}}+\mathbf{F}_{iw}^{\mathrm{shear}}$). In
order to produce the profiles shown below, $W_{in}$ is averaged in
steady states over small square bins of the size of less than
$\sigma/2$ and a time interval $\Delta t\sim 10$~s.

\subsection{Event driven hard sphere simulations of random driving}
\label{sec:edmd}
We simulate frictionless, smooth hard spheres using event driven
simulations \cite{poeschel05}.  We implemented a reduced collision
list version of the event driven algorithm in which the simulation box
is divided into cells and only collisions between pairs of particles
in neighboring cells are considered.  We do not employ sophisticated
data structures for the schedule of events and simply keep an ordered
list of collisions and cell boundary crossings.

When either the interparticle coefficient of restitution $e$ or
particle-wall coefficient of restitution $e_\mathrm{wall}$ is less
than unity, energy must be supplied to the system to balance the
energy dissipated in collisions.  We consider three ways of injecting
energy, which we term global kicking, local kicking and hot-wall. 

When the ``global kicking'' is in use, at every particle
collision we select a random particle pair distinct from the colliding
pair and give the selected particles equal and opposite velocity kicks drawn
from a normal distribution with zero mean.  The ``local kicking''
method draws pairs of particles from neighboring cell.  This method
may reduce the long wavelength velocity correlations that arise from
random pair kicking \cite{fiege09}.   In the vibrating system, the energy injection occurs only when
particles collide with the top and bottom boundaries.  To mimic this
feature of the vibrated system, we implemented what we call the
``hot-wall'' driving.  A kick in the $X$--$Y$ (``horizontal'') plane is administered to
a particle which collides with a confining plate.  The magnitude of
the kick is saved and a kick equal in magnitude and opposite in direction is administered to
the next particle which undergoes a collision with a confining plate.
The pairwise administration of kicks assures that the center of mass
momentum does not grow without bound \cite{bizon99}.

To study the importance of inelastic collisions on ordering we also
implemented a ``global energy sink'' with elastic ball-ball and
ball-wall collisions.  The energy injected by the random kicks is
removed by rescaling particle velocities by a constant factor after
every kick to keep the total kinetic energy constant.  Because all
velocities are rescaled by the same factor, all events in the list are
still valid modulo an appropriate rescaling of the event times.  The
standard deviation of the kicking velocity distribution, $v_{kick}$, relative to the fixed total
kinetic energy, combined with the number of kicks per collision,
determines how different the system is from the elastic system.  

To test for the appearance of the phase consisting of two layers with
square symmetry, we calculate the order parameter, $\Psi_8$, which measures
the degree of square order in the $X$--$Y$ plane.  It is defined as
\begin{equation}
  \label{eq:1}
  \Psi_8 = \frac{1}{N} \left| \sum_{i=1}^N \psi^8_{i} \right|, \quad
  \psi^8_i = \frac{1}{n_i} \sum_{\langle j\rangle} e^{8\pi I \theta_{ij}},
\end{equation}
where $I = \sqrt{-1}$, the sum over $\langle j\rangle$ in the
definition of $\psi^8_i$ is over $n_i$ neighbors of $i$ defined as
those spheres whose center is within a cutoff distance $1.35\sigma$ in
the $X$--$Y$ plane from the center of $i$ and $\theta_{ij}$ is the
angle between the $X$--$Y$ projection of $\mathbf{r}_i - \mathbf{r}_j$
and the $X$ axis.  When the system is perfectly ordered into a square
lattice in the $X$--$Y$ plane, $\Psi_8 = 1$, whereas a randomly
positioned collection of spheres has $\Psi_8 \sim 1/\sqrt{N}$.

\section{Vibrated layers}
\label{vibrated}
The simulations described in this paper were motivated by a series of
experimental results on a system consisting of a collection of
identical spherical particles on a horizontal plate vibrating
vertically, with a confining lid \cite{prevost04,melby05,reyes08}.
This system is well suited to investigate many of the complex
non-equilibrium effects observed in excited granular media and is
simple enough that it can be accurately modeled in molecular dynamics
simulations \cite{prevost04}.

\subsection{Phase Diagram}
The vibrated system shows a wide variety of steady states, depending
on the number of particles, the vibration frequency and amplitude, and
the gap between the plate and the confining lid
\cite{prevost04,melby05,reyes08,clerc08}.  For the gap spacing studied
here (1.75 $\sigma$), the spheres spontaneously organize into a
two-layer ordered phase showing square symmetry.  The ordered phase
forms at sufficiently high density, and is observed for a range of
vibration amplitudes and frequencies.  At slightly larger gaps, the
spheres form a two layer hexagonally ordered phase, while at smaller
gaps a variety of complex phases are observed \cite{melby05}.  This
surprising behavior can be understood, at least in part, by
considering the very similar behavior observed in equilibrium hard
sphere colloidal suspensions confined in a similarly narrow gap
\cite{pansu83}.  These systems show a complex phase diagram that
depends on the volume fraction of the colloidal particles and the
ratio of the confining gap to the particle diameter
\cite{pansu83,pieranski83,pansu84,schmidt96,schmidt97,zangi00}.  The
phase diagram can be explained in terms of entropy maximization.  The
configuration that gives the particles the most room to rattle around,
within the constraints imposed by the gap, will minimize the free
energy of the system. It remains unclear, however, whether this
picture can be extended to explain the non-equilibrium steady states
observed in the granular layer.  In the next two sections we describe
purely non-equilibrium effects that are not present in the colloidal
suspensions.

\subsection{Phase dependence of energy injection}
For a range of densities, the two-layer square phase coexists with a
disordered liquid-like phase, with a sharp boundary between the
phases. The square phase has a much higher density (nearly close
packed) and much lower granular temperature than the surrounding
liquid \cite{prevost04}.  As described in Section \ref{random}, these
effects are not present in systems of elastic spheres or of randomly
driven inelastic spheres, although the square phase is observed at
high densities.  One important feature of energy injection by
vibration that is not present in many idealized models of granular
forcing is that the rate of energy injection depends on the local
dynamics of the system.  We performed molecular dynamics simulations
to directly measure the rate of energy flow from the vibrating plate
to the two coexisting phases.

\begin{figure}[h]
\center
\includegraphics[width=12.5cm]{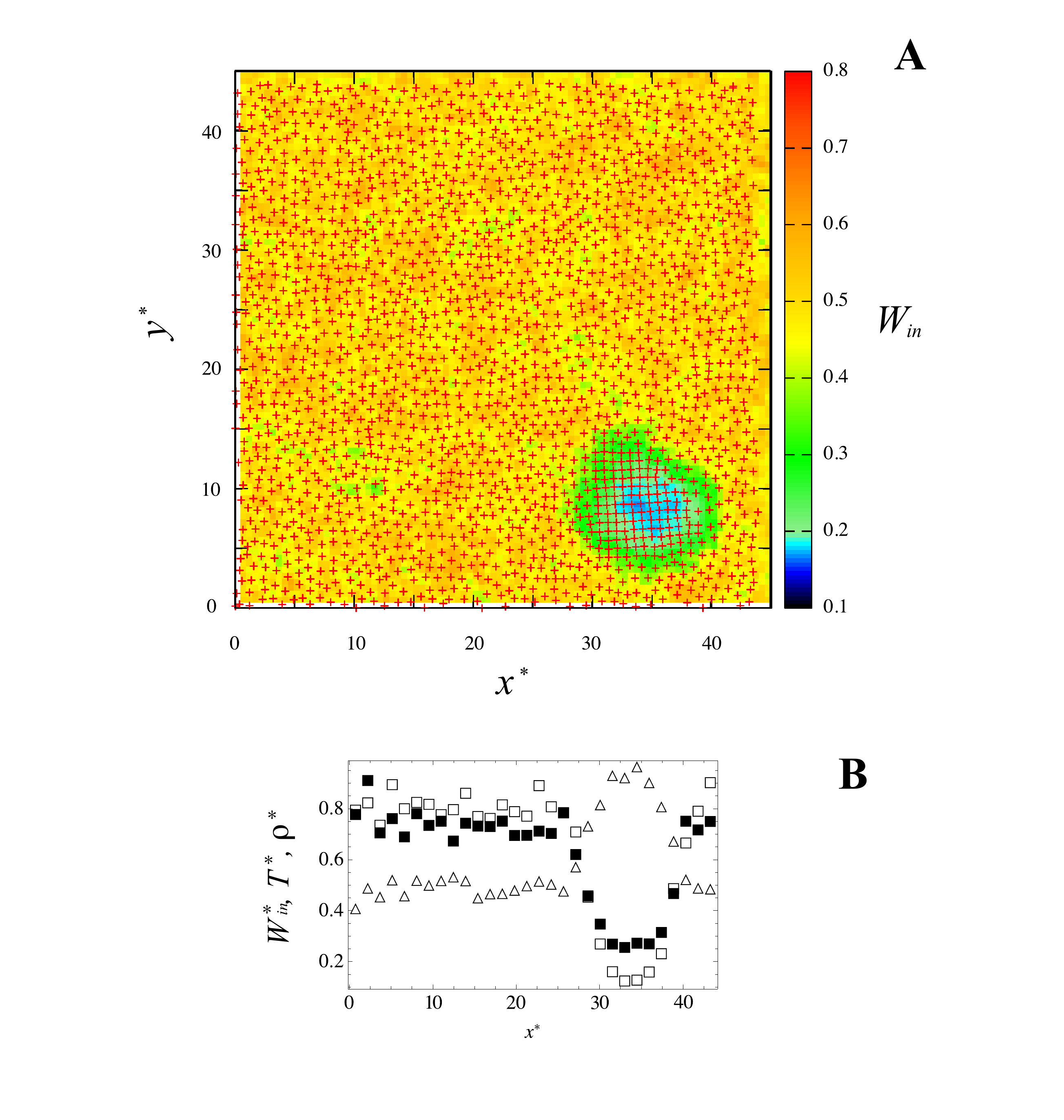}
\caption{{Energy injection in vibrated system is phase-dependent.  \small {\bf A.} The color map shows the square of the energy input, $W _{in}$, as defined in Eq. \ref{workinput}, with $x^*, y^*=x/\sigma, y/\sigma$. The position of the center of each ball at the midpoint of the averaging period is marked with $+$. The dramatic reduction in energy input in the lower right corner coincides with the ordered phase. {\bf B.} Normalized profiles of energy input $W_{in}^*$ ($\blacksquare$), horizontal granular temperature $T_h^*$ ($\square$) and  density $\rho^*$ ($\vartriangle$). The normalization is calculated according to $a^*(x)=a(x)/a_{max}$, where $a_{max}$ 
is the maximum value of $a$ in the system. The profiles are measured along a straight line passing through the center of the ordered phase. (N=2000, system size $L=44~\sigma$, gap spacing $h=1.75~\sigma$ ($\sigma=1.19063$~mm) vibration frequency $\nu=60$~Hz, input acceleration $\Gamma=A(2\pi \nu)^2/g=3.00$ ($g=9.80$~ms$^{-2}$). Plots result from averaging over 10 s.) }}
\label{input}
\end{figure}

Fig \ref{input}a shows a color map of the local average rate of energy
input.  Superimposed on the colormap are crosses representing the
instantaneous position of the spheres.  The square phase is clearly
visible, and exactly coincides with a region of dramatically reduced
energy input.  In order to more clearly show the correspondence, we
show in Fig. \ref{input}b the normalized energy input rate (solid
squares), granular temperature (open squares), and density
(triangles), calculated for a narrow strip that runs through the
ordered phase.  The implications of these results are discussed in
Section \ref{discussion}.

\subsection{Effects of inelasticity}
We recently reported the existence of a melting transition of the
square ordered phase as the vibration amplitude is increased, both in
experiments and computer simulations \cite{reyes08}.  Furthermore,
experiments showed that the melting of the crystalline phase occurs
significantly earlier in brass spheres than in less inelastic
stainless steel spheres, and computer simulations showed that the
ordered phase is not present at any vibration amplitude when the
inelasticity is large.  A reduction in the liquid-solid coexistence
region at fixed vibration amplitude was also observed in computer
simulations in \cite{clerc08}.  These results indicate that ordering
is suppressed by inelasticity and that the strength of dissipation can
qualitatively alter the phase diagram of even the simplest granular
media. One of the motivations for the work described here is to see if
that effect is unique to vibrated granular systems.

\section{Randomly driven layers}
\label{random}
In order to elucidate the effects of forcing and dissipation on the
phase diagram, we performed simulations on randomly driven hard
spheres.  Although there is no experimental system that closely
matches the random driving, it has a number of advantages relative to
the vibrated system.  For a given geometry, the state of the system is
fully specified by density and inelasticty, 
whereas amplitude and
frequency are also relevant parameters for the vibrated system.  The
behavior of the randomly driven hard spheres can be compared with
equilibrium hard sphere suspensions by observing the behavior as $e$
approaches unity.

\subsection{Elastic hard spheres}

\begin{figure}
  \center
   \includegraphics[width=6cm]{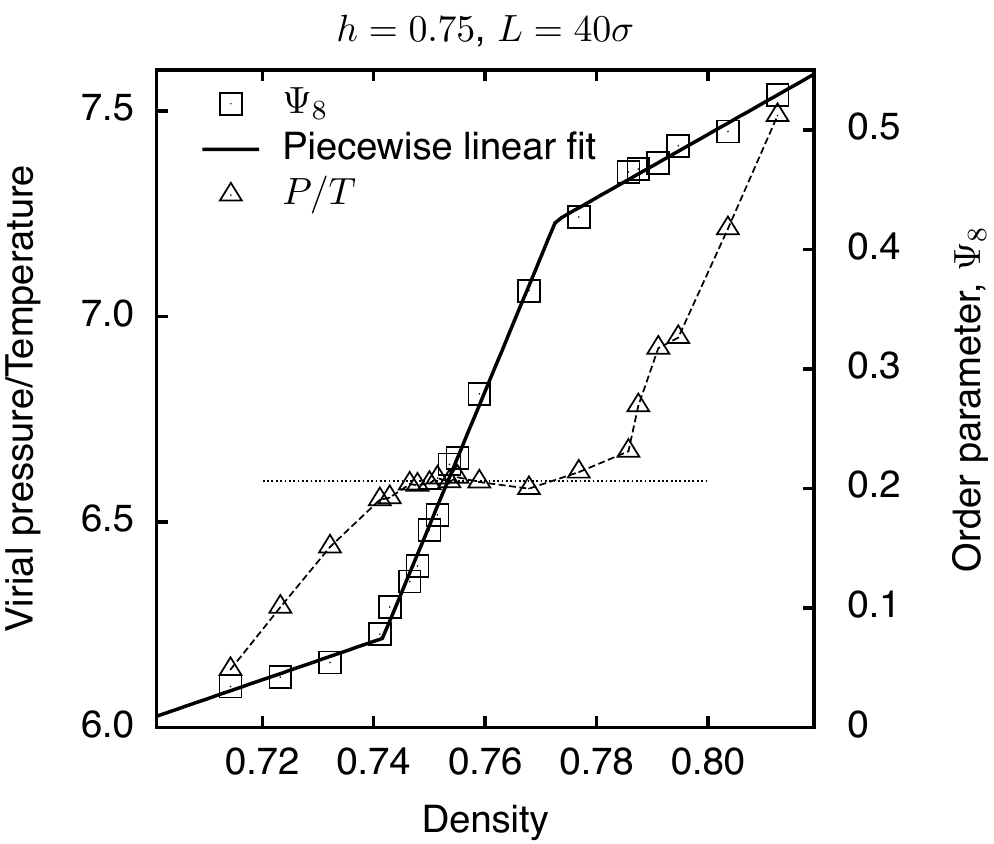}
  \caption{Virial pressure scaled by the temperature (triangles)
    measured in a series of constant volume simulations of elastic
    spheres shows a plateau as a function of density indicative of a
    coexistence between two phases.  The square order parameter
    $\Psi_8$ (squares) exhibits a rapid growth rate in the same
    density window.  A piece-wise linear fit to $\Psi_8$ yields the
    densities bracketing the coexistence region.}
  \label{elast}
\end{figure}

Elastic hard spheres confined to a gap of 1.75 $\sigma$ undergo a 
transition from 
disorder (liquid) to square order (crystal) upon
increasing density
\cite{pansu83,pieranski83,pansu84,schmidt96,schmidt97,zangi00}.  At
constant volume, the two phases coexist in a range of densities.  The
signature of 
coexistence is a plateau in the pressure as a 
function of
density.  Figure \ref{elast} shows the virial pressure scaled by the
temperature (which is constant in an elastic, constant volume system)
as a function of density, defined as $AH/(N\sigma^3)$, where $A=L^2$
is the area of the confining plates.  The plateau is indicative of
coexistence between the densities of roughly 0.75 and 0.77.

Another signature of the phase coexistence is found by examining the
variation of the order parameter $\Psi_8$ with density.  In the
coexistence region, the system presumably separates into two distinct
regions characterized by different order parameters.  Since the
characteristic densities of the phases should remain the same in the
coexistence region, the fraction of the system in the ordered state
grows linearly from 0 to 1 as the density increases.  The mean order
parameter averaged over the whole system should therefore exhibit a
linear increase as well.  As shown in Figure \ref{elast}, an estimate
of the boundaries of the coexistence region determined from piecewise linear fits to the  order parameter vs. density curve is roughly consistent
with the extent of the plateau in the pressure vs.~ density graph.

\begin{figure}
  \centering
\includegraphics[width=6cm]{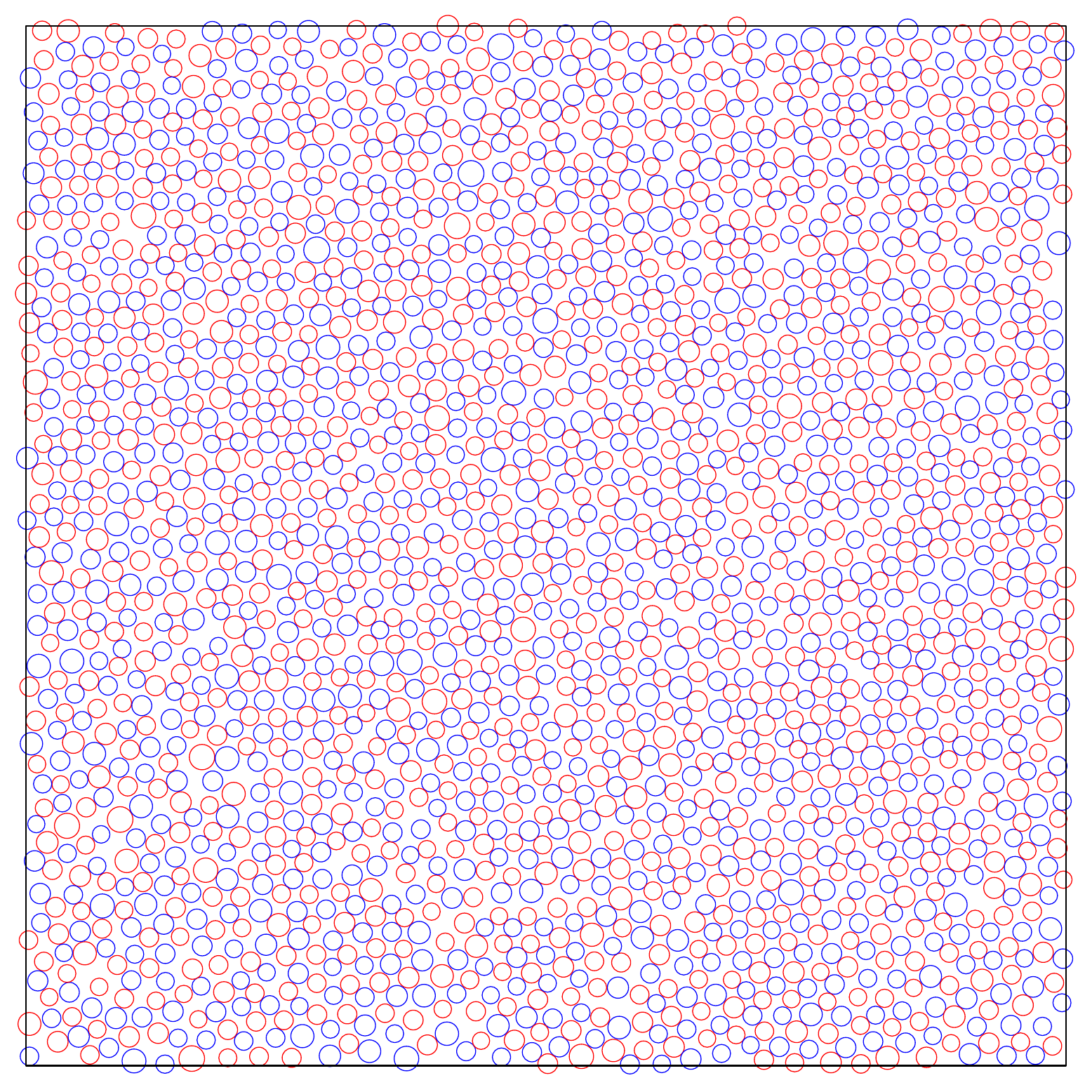}
  \caption{Snapshot of the $z=H/2$ section through the elastic system
    in the coexistence region.  The density is $\rho = 0.768$, $L =
    40\sigma$ and $h = 1.75\sigma$.  The larger circles result from
    spheres with centers closer to the $z = H/2$ plane, and the color
    indicates whether the center of a particular sphere is below
    (blue) or above (red) that plane.}
  \label{coex}
\end{figure}

A snapshot of the elastic system at a density that is in the middle of
the coexistence region is shown in Fig. \ref{coex}.  Unlike the
vibrated inelastic system, there is no sharp separation between the
ordered and disordered phases (compare with Fig. \ref{input}).  We
have looked extensively for indications of phase separation, including
very long simulation runs and very large systems, but have always
found a high degree of intermixing.  Note that because this system is
quasi-2D, it is perhaps not surprising that fluctuations dominate over
what is presumably a very small effective surface tension.  In the
vibrated inelastic system, by contrast, the phase separation is
dramatic, nearly instantaneous, and can be observed in both small and
large systems \cite{reyes08}.

\subsection{Effects of Inelasticity}
\begin{figure}
  \center
\includegraphics[width=12cm]{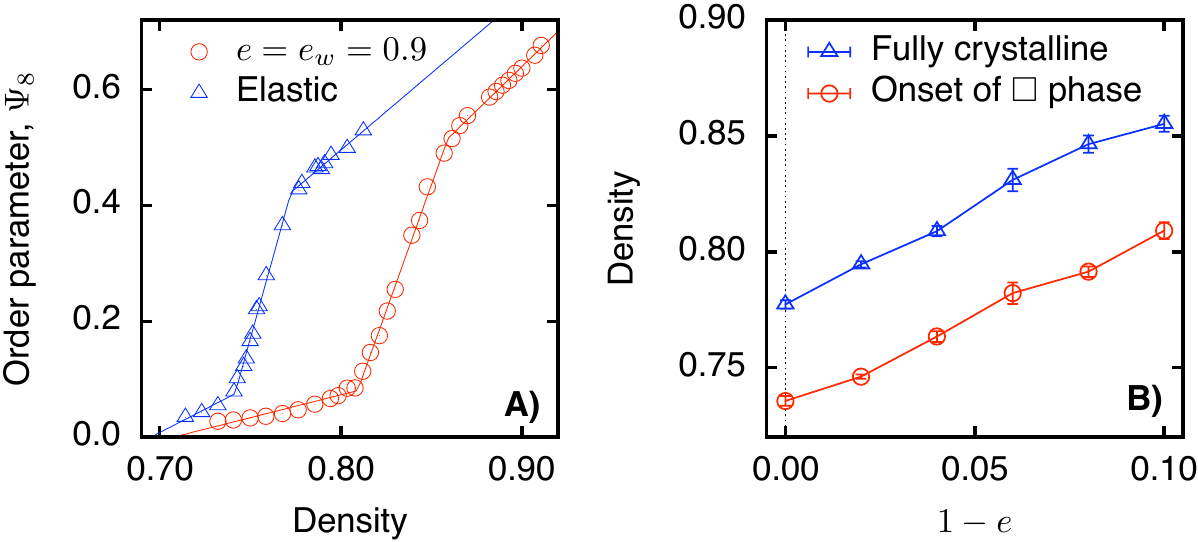}
  \caption{\textbf{A)} The square order parameter in an elastic system
    compared to that of an inelastic system driven by random pair
    kicking (the global kicking method).  The extracted boundaries of
    the coexistence region are plotted in panel \textbf{B)} as a
    function of sphere inelasticity $e$ for a globally kicked system
    (wall inelasticity $e_\mathrm{wall} = e)$.}
  \label{inelast}
\end{figure}

When particles are inelastic, energy must be continuously supplied to
the system to maintain a dynamic steady state.  We first focus on
systems in which energy is supplied via the ``global kicking''
mechanism described in \ref{sec:edmd}.  Fig. \ref{inelast}A shows the
order parameter vs.~density for spheres with
$e=e_\mathrm{wall}=0.9$.  The data for elastic spheres (also shown in
Fig.~\ref{elast}) is presented for comparison.  As for the vibrated
spheres, the transition to the ordered phase is shifted to a higher
density in the inelastic system.  The evolution of the transition region, determined piecewise linear fits to the  order parameter vs. density curves, appears to be a
smooth function of inelasticity, as shown in Fig. \ref{inelast}B.
This suggests that the effect of the inelasticity can be thought of as
a perturbation to the phase behavior of the elastic system.  However,
to our knowledge, there is no appropriate perturbation theory for the equilibrium statistical mechanical explanation of
the transition to a state of square symmetry.

Images of the coexistence region in the inelastic system are
indistinguishable from the elastic system.  In particular, phase
segregation between the ordered and disordered phases is never
observed (unlike the vibrated system).  Quantitative analyses of density-density and order
parameter correlation functions show no features that clearly reveal
effects of inelasticity, although our sensitivity is limited by the
relatively small systems studied here. 

Unlike the elastic system, plots of $P$ or $P/T$ vs.~density do not,
in general, show a plateau in the coexistence region (see
Fig.~\ref{elast}).  In the inelastic system, the steady state granular
temperature is determined by the balance between forcing and
dissipation.  Thus the temperature will change if the configuration of
the system changes, even if all parameters are kept constant.  For a
homogeneous hard sphere system, $P/T$ should be independent of
temperature, but we do not find a coexistence plateau in $P/T$.  We
hypothesize that this is because the temperature is non-uniform in the
inelastic system.  Even though the energy injection is homogeneous,
denser regions will dissipate energy more quickly.  Thus we expect
that the ordered phases will be cooler, although the difference is
more subtle than in the vibrated system and does not seem to yield a
readily observable correlation between the temperature and order
parameter.

\section{Effect of forcing and dissipation}
\begin{figure}
  \center
  \includegraphics[width=12cm]{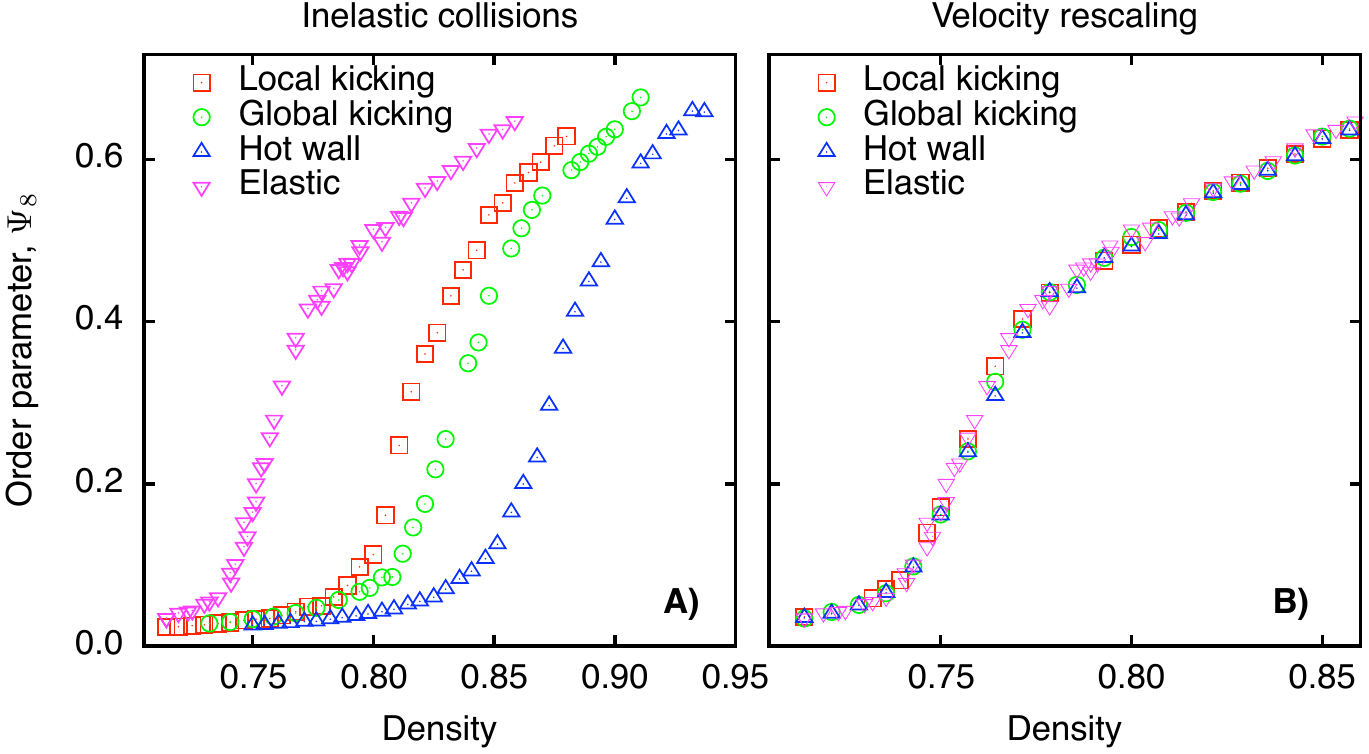}
  \caption{The effect of different forms of energy injection and dissipation on the square ordering
    of the confined spheres.  \textbf{A)} Order parameter vs. density
    for different forms of energy injection ($e=e_\mathrm{wall}=0.9$).
    \textbf{B)} Order parameter vs. density for global dissipation
    (velocity rescaling), with either local or global random kicking
    or a hot wall ($v_{kick}= 0.1*\sqrt{2T}$).}
  \label{lotsocurves}
\end{figure}

The results from the vibrated and randomly driven systems show that
inelasticity suppresses the ordering transition, but the mechanism for
the suppression is not well understood.  In order to distinguish the
relative importance of energy injection vs. energy dissipation, we
performed simulations with different methods for energy input and
energy dissipation.  The results, summarized in Figure
\ref{lotsocurves}, are discussed below.

The pair kicking injection methods produce long-lived velocity
correlations. (Ref.~\cite{fiege09} studies the correlations that arise
as a result of the ``global kicking'' method.)  Picking the kicked
pair from neighboring cells reduces, but does not altogether eliminate,
the velocity correlations.  Although we have not conducted a careful
study of the ``hot wall'' driving, it too is likely to introduce long
range velocity correlations, albeit of a different nature.
Fig. \ref{lotsocurves}A shows that for the same inelasticity ``hot
wall'' driving ($\bigtriangleup$) suppresses square ordering the most,
while the suppression for global ($\odot$) and local ($\boxdot$) kicking are comparable, with the local kicking being the least effective at
suppressing square order.

We employed the velocity rescaling ``global energy sink'' described in
\ref{sec:edmd} to help to separate the effects of forcing from the effects of 
dissipation.  Whereas collisions are
elastic, the particles are given random kicks according to each of the
three driving schemes, and the injected energy is removed ``globally''.  The order
parameter-density curves, shown in Fig. \ref{lotsocurves}B, are
indistinguishable from the elastic case.  This surprising result suggests that velocity
correlations or other effects due to driving alone are not sufficient to suppress
ordering. 

In summary, we find that inelastic collisions suppress the transition
to the ordered phase, and that the suppression is approximately
proportional to the degree of inelasticity, but the exact amount of the suppression depends on the form of energy injection.  In contrast, random
kicking without inelastic collisions does not appear to significantly affect  the location of the ordering
transition.

\section{Discussion}
\label{discussion}
The results described in this paper provide several insights into the
differences between phase transitions in excited granular layers and
their equilibrium analogs, but also point to several unanswered
questions.

Here we have shown directly that the energy flow from the vibrating
plate into the dense ordered phase is much lower than into the
surrounding dilute liquid.  It has been shown previously that
velocity-dependent driving can lead to clustering in driven granular
monolayers \cite{cafiero00}, and is presumably the mechanism
responsible for the 'collapse' observed in monolayers at small
vibration amplitude \cite{olafsen98}.  Unlike the collapse, however,
the ordered phases described in this paper are stable at high
vibration amplitudes and occur at densities where elastic systems also
show ordered phases. The absence of a clear phase separation between
the ordered and disordered phases in confined elastic or randomly
driven spheres suggests that the dramatic separation seen in the
vibrated system can be traced to the inhomogeneous heating.

Our simulations of various forms of randomly driven hard spheres
show that the suppression of the ordering transition upon
increased inelasticity seen in vibrated layers is fairly generic, and
that, at least for hard spheres, the transition evolves smoothly to
the elastic limit as the inelasticity goes to zero.  The significance
of this result is that the phase transition in the weakly inelastic
system should be explicable using the same machinery as the transition
in the elastic case, which is driven by entropy maximization through
geometric packing effects.  However, non-equilibrium effects need to
be included to explain how the phase boundaries shift with increasing
dissipation.

The results from the simulations of different forms of random energy
injection suggest a complex interplay between energy injection, dissipation, and ordering.   The fact that the details
of the forcing and dissipation will modify the location of the phase
boundaries is perhaps not surprising.  For example, the velocity correlations induced by the forcing could change the pressure in the disordered phase, which would likely shift
the van der Waals instability that leads to the formation of the
higher density phase. In this context the absence of any measurable shifts in the location of the ordering transition in simulations with random kicking and a global energy sink is perplexing.

While we have focused on somewhat idealized model systems, the effects
are potentially relevant to a wide class of excited granular media.
For most forms of energy injection, such as gas fluidization or shear
flow, the local rate of heating will depend on the local state of the
granular media, as we have observed with the vibrated
layers. The transition regime between a freely flowing
granular liquid and rigid granular packing is critical in many
situations, and the final steady state of a forced granular system
will in general depend on the details of the balance between
energy input and collisional cooling.  A better understanding of the
effects of forcing and dissipation on the transition in well
controlled model systems will hopefully assist with the development of
robust models for more complex systems.

\section{Acknowledgments}
This work was supported by NASA under award number NNC04GA63G.  One of
the authors (F. V. R.) acknowledges also support from Ministerio de
Ciencia e Innovaci\'on (Spain) through contract No. FIS2007-60977 and
Junta de Extremadura (Spain), through contract No. GRU09038.

\end{document}